\documentstyle[prl,aps,epsf,multicol]{revtex}
\begin{document}
\draft

\title{High Temperature Conductance of the Single Electron Transistor}
\author{Georg G\"oppert and Hermann Grabert}
\address{Fakult\"at f\"ur Physik, Albert--Ludwigs--Universit{\"a}t, \\
Hermann--Herder--Stra{\ss}e~3, D--79104 Freiburg, Germany}

\date{\today}
\maketitle
\widetext

\begin{abstract}
The linear conductance of the single electron transistor is
determined in the high temperature limit. Electron tunneling is 
treated nonperturbatively by means of a path
integral formulation and the conductance is obtained from Kubo's 
formula. The theoretical predictions are valid for arbitrary conductance
and found to explain recent experimental data.
\end{abstract}


\raggedcolumns
\begin{multicols}{2}
\narrowtext 

Considering applications of single electron tunneling, one is faced
with a dilemma. On the one hand, the conventional approach \cite{Nato}
consider systems with small tunneling conductance
$G_T \ll G_K=e^2/h$ implying nearly perfect charge
quantization. On the other hand, when using these devices, e.g.\ as highly
sensitive electrometers \cite{Fulton} or for thermometry
\cite{Pekola94}, a large current signal is desirable meaning large
tunneling conductance. Since for $G_T \gg G_K$ charging effects
disappear, a compromise must be found in practice. This problem has
spurred considerable interest in the precise behavior of single charge
tunneling devices for large conductance \cite{strongtunneling,WangSluggon}.

\begin{figure}
\begin{center}
\leavevmode
\epsfxsize=0.4 \textwidth
\epsfbox{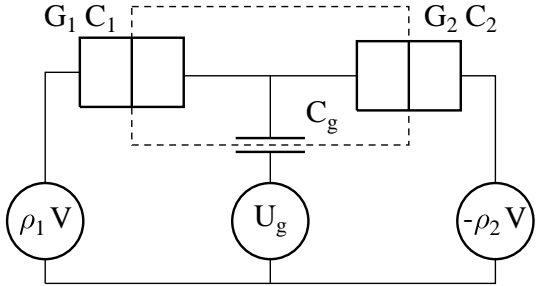}
\end{center}
\caption{Circuit diagram of the single electron transistor.}
\label{fig:fig1}
\end{figure}

In this letter, we consider the single electron transistor
(SET) at
high temperatures for {\it arbitrary} tunneling conductance of the
junctions. The SET consists of
two tunnel junctions in series, with tunneling
conductance $G_1$ and $G_2$ and capacitance $C_1$ and $C_2$,
respectively, biased
by a voltage source $V$ which may be split among the left and right
branches into $\rho_1 V$ and
$\rho_2 V$ with $\rho_1+\rho_2=1$, cf.\ Fig.\ \ref{fig:fig1}. 
The island in between the junctions is
connected via a gate capacitance $C_g$ to a control voltage $U_g$
shifting the electrostatic energy of the system continuously. The
important energy scale is the charging energy $E_c=e^2/2C$ with the
island capacitance $C=C_1+C_2+C_g$. For weak electron tunneling, $E_c$ is
the energy needed to charge
the island with one excess electron at vanishing gate voltage $U_g=0$. Due to the periodicity
of the Hamiltonian in $U_g$, the conductance is a periodic function
with period 1 of
the dimensionless voltage $n_g=U_g C_g/e$ \cite{Averin}. 
Specifically, we are
interested in the linear 
dc conductance for small transport voltage, that may be calculated from
the Kubo formula
\begin{equation}
\!G = \lim_{\omega \rightarrow 0} \frac{1}{\hbar \omega} {\rm Im}
    \lim_{i \nu_l \rightarrow \omega + i\delta}
    \int_0^{\hbar \beta} \! \! \! d\tau e^{i\nu_l \tau}
    \langle I^{(1)}(\tau) I^{(2)}(0) \rangle
 . 
\label{eq:kubo1}
\end{equation}
Since the dc current through both junctions coincides, the first
current operator $I^{(1)}$ may be an arbitrary linear
combination $I^{(1)}=\epsilon_1
I_1-\epsilon_2 I_2$ (with $\epsilon_1+\epsilon_2=1$) of the current
operators $I_1$ and $I_2$ through junctions
$1$ and $2$, respectively. The relative "$-$" sign
comes from the
opposite directions of $I_1$ and $I_2$, which are both positive for flux
onto the island. The second current operator $I^{(2)}=\kappa_1 I_1
-\kappa_2 I_2$, with the relative weights $\kappa_1=(C_2+\rho_1C_g)/C$ and
$\kappa_2=(C_1+\rho_2C_g)/C$, is determined
by linear response theory from the coupling of the transport voltage $V$. To
evaluate the
current-current correlator we employ a generating
functional
\begin{equation}
\! Z[\xi_1,\xi_2]={\rm tr} \, T_\tau \exp \!\!
 \left\{
  \!-\frac{1}{\hbar}\int_0^{\hbar \beta} \!\!\! d\tau 
  \bigg[ H -\! \sum_{i=1,2} \! I_i\xi_i(\tau) \bigg] \!
 \right\},
\end{equation}
where $H$ is the Hamiltonian of the system for $V=0$, and $T_\tau$ is
the time ordering operator for imaginary times $\tau$. The
correlator is then given by a second order variational derivative
relative to the auxiliary fields $\xi_1$ and $\xi_2$. In the phase
representation \cite{SchoenRep}, we get for the generating functional
\begin{equation}
Z[\xi_1,\xi_2]=\int D[\varphi] \exp
 \left\{
  -\frac{1}{\hbar} S[\varphi,\xi_1,\xi_2]
 \right\},
\end{equation}
with the effective action
\begin{equation}
S[\varphi,\xi_1,\xi_2]=S_c[\varphi]+S_1[\varphi,\xi_1]+S_2[\varphi,\xi_2].
\end{equation}
The first term on the rhs is the Coulomb action
\begin{equation}
S_c[\varphi]=\int_0^{\hbar \beta}d\tau
 \left[
  \frac{{\dot \varphi}^2(\tau)}{4 E_c} - in_{\rm g}{\dot \varphi}(\tau)
 \right]
\end{equation}
describing the Coulomb charging of the island in presence of an
applied gate voltage, and the effective tunneling actions $(i=1,2)$
\begin{eqnarray}
S_i[\varphi,\xi_i] 
&=& g_i \int_0^{\hbar \beta} d\tau d\tau' 
 \alpha(\tau-\tau')       \nonumber  \\
&&
(1-ie\xi_i(\tau))(1+ie\xi_i(\tau'))
 e^{i[\varphi(\tau)-\varphi(\tau')]}
\end{eqnarray}
describe quasi-particle tunneling through junctions $1$ and $2$,
respectively. Here $g_i=G_i/G_K$ is the dimensionless conductance of
junction $i$, and the kernel $\alpha(\tau)$ may be
written as Fourier series
\begin{equation}
\alpha(\tau)=-\frac{1}{4\pi \beta}\sum_{n=-\infty}^{\infty}
 |\nu_n| \, e^{-i\nu_n \tau},
\end{equation}
where the $\nu_n=2\pi n/\hbar\beta$ are Matsubara frequencies.
For vanishing auxiliary fields the action reduces to the action of
the single electron box \cite{WangSluggon} and thus the generating
functional reduces to the box partition function
$Z=Z[0,0]$. Performing the variational derivatives explicitly, we get
for the correlator
\begin{equation}
\hspace{-0.2cm}
\langle
 I_j(\tau) I_{j'}(\tau')
\rangle
=
\langle 
 I_j(\tau) I_j(\tau')
\rangle^E \delta_{j,j'} +
\langle 
 I_j(\tau) I_{j'}(\tau')
\rangle^F.
\label{eq:corr0}
\end{equation}
Since the auxiliary fields are in the argument of an exponential, there
are two contributions. The first term comes from the second order
variational derivative of the action and reads 
\begin{eqnarray}
\langle 
 I_j(\tau) I_j(\tau')
\rangle^E
&=&
\frac{2e^2}{\hbar} g_j \alpha(\tau-\tau') 
\frac{1}{Z}
\int D[\varphi]       \nonumber   \\
&&
\exp\left\{-\frac{1}{\hbar}S[\varphi]\right\}
 \cos[\varphi(\tau)-\varphi(\tau')],
\label{eq:corr1}
\end{eqnarray} 
with the box action $S[\varphi]=S[\varphi,\xi_1=0,\xi_2=0]$. The
second term in Eq.\ $(\ref{eq:corr0})$ involves a multiplication of two current
functionals arising as first order variational derivatives of the action
\begin{eqnarray}
\langle 
 I_j(\tau) I_{j'}(\tau')
\rangle^F
&=&
g_j g_{j'} \frac{1}{Z}\int
D[\varphi]        \nonumber  \\
&&
\exp\left\{-\frac{1}{\hbar}S[\varphi]\right\}
I[\varphi,\tau] I[\varphi,\tau'],
\end{eqnarray}
with the current functional
\begin{equation}
I[\varphi,\tau]=\frac{2e}{\hbar}\int_0^{\hbar\beta}d\tau'
\alpha(\tau-\tau')\sin[\varphi(\tau)-\varphi(\tau')].
\end{equation}
Taking into account that 
$\langle 
 I_j(\tau) I_{j}(\tau')
\rangle^E/g_j$ and $\langle 
 I_j(\tau) I_{j'}(\tau')
\rangle^F/ g_j g_{j'}$ depend only on the dimensionless parallel
conductance $g=g_1+g_2$ and thus are
independent of the indices $j$ and $j'$, the conductance may be
written as 
\begin{eqnarray}
G
&=&
\epsilon_1\kappa_1 ( g_1 E + g_1^2 F)-
2\epsilon_1\kappa_2  g_1 g_2 F 
                       \nonumber      \\
&&
+ \epsilon_2\kappa_2 ( g_2 E + g_2^2 F), 
\end{eqnarray}
where
\begin{equation}
\! E=\lim_{\omega \rightarrow 0} \frac{1}{\hbar \omega} {\rm Im}
    \lim_{i \nu_l \rightarrow \omega + i\delta}
    \int_0^{\hbar \beta} \! d\tau \, e^{i\nu_l \tau}
    \frac{   
      \langle I_1(\tau) I_1(0) \rangle^E 
         }{g_1}
\end{equation}
and\begin{equation}
\! F=\lim_{\omega \rightarrow 0} \frac{1}{\hbar \omega} {\rm Im}
    \lim_{i \nu_l \rightarrow \omega + i\delta}
    \int_0^{\hbar \beta} \! d\tau \, e^{i\nu_l \tau}
    \frac{
       \langle I_1(\tau) I_1(0) \rangle^F 
         }{g_1^2}
\end{equation}
Since the conductance does not depend on
the specific choice of the parameters $\epsilon_i$ and $\rho_i$, we
then find that
\begin{equation}
G=E/(G_1^{-1}+G_2^{-1}),
\label{eq:kubo2}
\end{equation}
This is a formally exact expression for the linear dc conductance. 

To proceed, we make explicit the sum over winding numbers $k$ of the
phase and write the correlator (\ref{eq:corr1}) in the form
\begin{eqnarray}
\langle
 I_1(\tau) I_1(0)
\rangle^E /g_1
&=&
 \frac{2e^2}{\hbar} \alpha(\tau) \frac{1}{Z}\sum_{k=-\infty}^{\infty}
\int\limits_{\varphi(0)=0}^{\varphi(\hbar\beta)=2\pi k}
 D[\varphi]                  \nonumber  \\
&&
\hspace{-0.3cm}
\exp\left\{ -\frac{1}{\hbar}S[\varphi] \right\} 
 \cos[\varphi(\tau)-\varphi(0)]. 
\label{eq:korr1}
\end{eqnarray}
For given winding number $k$, the path integral may be evaluated
approximately by expanding about the classical path ${\bar
\varphi}(\tau)=\varphi(0)+\nu_k \tau$. An arbitrary path of winding
number $k$ is of the form
$\varphi(\tau)={\bar \varphi}(\tau)+\zeta(\tau)$ with
$\zeta(0)=\zeta(\hbar \beta)=0$, and the action can be written in terms
of the Fourier coefficients $\zeta_n=\zeta_n'+i \zeta_n''$ as
\begin{equation}
\!\!\! S[{\bar \varphi}+\zeta]
=
- 2 \pi i k n_g + S_{\rm cl}^{(k)}+ \delta^2S^{(k)}+
\delta^4S^{(k)}+ ... \,.
\label{eq:action}
\end{equation}
Here
\begin{equation}
S_{\rm cl}^{(k)}=\frac{\pi^2 k^2}{\beta E_c}+|k|\frac{g}{2},
\end{equation}
is the action of the classical path. The second term reads
\begin{equation}
\delta^2S^{(k)}=\sum_{n=1}^\infty \lambda_n^{(k)}
\left( \zeta_n'^2+\zeta_n''^2 \right),
\end{equation}
with the eigenvalues
\begin{equation}
\lambda_n^{(k)}=\frac{2\pi^2n^2}{\beta E_c}+g \Theta(n-|k|)(n-|k|).
\end{equation}
The term $\delta^4S^{(k)}$ is of fourth order in the Fourier
components $\zeta_n$ and not given explicitly here. 
Since the $\lambda_n^{(k)}$ are large for small $\beta E_c$, the
expansion $(\ref{eq:action})$ about the classical path converges rapidly
for high temperatures. 

Rewriting the
cosine function in Eq.\ $(\ref{eq:korr1})$ as a sum of exponentials,
we get for the correlator the expansion
\end{multicols}
\widetext
\begin{equation}
\langle
 I_1(\tau) I_1(0)
\rangle^E /g_1
= \frac{2 e^2}{\hbar} \alpha(\tau) \frac{1}{Z} \sum_{k=-\infty}^{\infty}
 C_k e^{2\pi i k n_g} \cos(\nu_k \tau) 
 \exp\left[-2\sum_{n=1}^\infty 
 \frac{1-\cos(\nu_n \tau)}{\lambda_n^{(k)}}\right] 
 \left[1+S_4^{(k)}+... \right],
\label{eq:korr2}
\end{equation}
\begin{multicols}{2}
\narrowtext
where the coefficients $C_k$ read
\begin{equation}
C_k=\frac{\Gamma(1+k_+)\Gamma(1+k_-)}{\Gamma^2(1+k)\Gamma(1+u)}
 e^{-S_{\rm cl}^{(k)}}, 
\end{equation}
with $k_\pm=k+\frac{u}{2}\pm\frac{1}{2}\sqrt{4uk+u^2}$ and  
$u=g \beta E_c/2\pi^2$. The dominant corrections to the
semiclassical approximation are described by
\begin{eqnarray}
S_4^{(k)}
&=&
 - \frac{1}{2} g \hbar \beta
   \sum\limits_{m,n\not=0}       
   \frac{1}{\lambda_m^{(k)} \lambda_n^{(k)}}
  \Big[
    {\tilde \alpha}(\nu_k)-2{\tilde \alpha}(\nu_{n+k})
              \nonumber  \\
& &
  \quad
    -2{\tilde \alpha}(\nu_{n-k})
    +{\tilde \alpha}(\nu_{m+n+k})+{\tilde \alpha}(\nu_{m+n-k})
  \Big].   
\label{eq:fourth}
\end{eqnarray}
The corresponding expansion of the partition function $Z$ reads 
\begin{equation}
Z=\sum_{k=-\infty}^\infty C_k e^{2\pi ik n_g}
\left[1+S_4^{(k)}+... \right],
\label{eq:partition}
\end{equation}
with the same correction 
$(\ref{eq:fourth})$. The expansions
$(\ref{eq:korr2})$ and $(\ref{eq:partition})$ proceed in powers of 
$\beta E_c$, however, terms involving $u=g\beta E_c/2\pi^2$ are kept 
to {\it all} orders. This ensures a meaningful result in the limit 
of moderately high temperatures also for large conductance $g$. 

From Eq. $(\ref{eq:korr2})$ one obtains for the Fourier coefficients
\begin{equation}
E(\nu_l)=\int_0^{\hbar \beta} d\tau \, e^{i\nu_l \tau}
    \langle I_1(\tau) I_1(0) \rangle^E /g_1
\end{equation}
the high temperature expansion
\end{multicols}
\widetext
\begin{eqnarray}
&
\! E(\nu_l) \! = \!
\frac{4\pi}{\hbar} \frac{1}{Z} \! \sum\limits_{k=-\infty}^\infty 
\!\! C_k e^{2\pi i kn_g} 
\exp\left[ \!-2\sum\limits_{n=1}^\infty \frac{1}{\lambda_n^{(k)}} \right]
\! \left\{
{\tilde \alpha}(\nu_{l+k})+  
\! \sum\limits_{m \not=0} \! \!
\frac{{\tilde \alpha}(\nu_{l+k+m})}{\lambda_m^{(k)}}+
\! \frac{1}{2} \! \sum\limits_{m,n \not=0} \!\!
\frac{{\tilde \alpha}(\nu_{l+k+m+n})}{\lambda_m^{(k)}\lambda_n^{(k)}}+
{\tilde \alpha}(\nu_{l+k})S_4^{(k)} 
\! + \! ...
\right\}\!.&
\label{eq:fourier}
\end{eqnarray}
\begin{multicols}{2}
\narrowtext
When these coefficients are analytically continued in the complex 
$\nu$ plane, they are analytic on each half plane 
Re$\,\nu{\tiny{\lower 2pt \hbox{$<$} \atop \raise 5pt \hbox{$>$}}}0$ 
with a cut along the imaginary axis \cite{Baym}. 
The representation of $E(\nu)$ as a sum over winding numbers $k$ shifts
this cut to Re$\,\nu=k$ for the $k\,$th term of the sum. Thus, in the 
phase representation, only the full sum shows the analytic properties
underlying the conductance
formula $(\ref{eq:kubo1})$. To deal with this problem we formally
change to the charge representation, 
perform the analytic continuation and the $\omega \rightarrow 0$
limit there, and then go back to the
phase representation.
This way the high temperature expansion of the conductance
$(\ref{eq:kubo2})$ may be evaluated to read
\begin{eqnarray}
G&=& [1/(G_1^{-1}+G_2^{-1})]  Z^{-1} 
  \exp\left\{-2[\gamma+\psi(1+u)]/g \right\} 
  \nonumber \\
&&
  \big\{
  1 - \psi'(1+u) (\beta E_c/ \pi^2)
  \nonumber \\
&&
 \hspace{0.5cm}
  + \left[ g \sigma(u)  + \tau(u)  \right]  
      (\beta E_c/ 2 \pi^2)^2
   +{\cal O}(\beta E_c)^3 \big\}.
\label{conductance}
\end{eqnarray}
The dependence on $u=g\beta E_c/2\pi^2$ is given in terms of the digamma 
function $\psi(z)$ and two auxiliary functions 
\begin{eqnarray}
\sigma(u)
&=&
  \frac{\gamma+\psi(1+u)-u\psi'(1+u)}{u^2}
  \nonumber \\
&&
 + \int_0^1 \! \mbox{d}v \, \frac{2v(1-v^u)\phi(v,1,1+u)}{(1-v)u}
\end{eqnarray}
and
\begin{eqnarray}
\tau(u) 
&=&
  \frac{\pi^2}{6u} \psi(1+u)
  -\frac{[\psi(1+u)+\gamma]^2}{u^2} 
  +\int_0^1 \! \mbox{d}v \, \Xi(u,v)
  \nonumber \\
&&
  -\frac{3\gamma\psi'(1+u)+\psi(1+u)[\frac{\pi^2}{6}+2\psi'(1+u)]}{u},
\end{eqnarray}
with Lerch's transcendent $\phi(z,u,v)$ \cite{Magnus} and
\end{multicols}
\widetext
\begin{eqnarray}
\Xi(u,v)&=&
\frac{\psi'(1+u)}{u \ln(v)}+
\frac{1}{(1-v)u}
\Bigg\{ 
  \frac{1-2v^u}{v}
  \left[ \ln(v)\phi(\frac{1}{v},1,1+u)+\phi(\frac{1}{v},2,1+u)\right]    
  +2v\phi(v,2,1+u) 
                                  \nonumber\\
&&                     
  -\frac{2(1-v^u)}{u}
  \left[\ln(1-v)+v\phi(v,1,1+u)\right]  
  +v^u\left[2\ln(v)\ln(1-v)
  +{1 \over 2} \ln^2(v)+3\mbox{Li}_2(1-v)\right]
\Bigg\}.
\end{eqnarray}
\begin{multicols}{2}
\narrowtext
The high temperature expansion of $Z$ is straightforward and reads
\begin{eqnarray}
Z
&=&
  1 + g \sigma(u) (\beta E_c/2\pi^2)^2  +{\cal O}(\beta E_c)^3
                                  \nonumber\\
&&                     
 \hspace{0.3cm}
  + 2 C_1 \cos(2 \pi n_g) \left[ 1+{\cal O}(\beta E_c)^2 \right],
\label{partition}
\end{eqnarray}
which combines with Eq.\ $(\ref{conductance})$ to yield an analytical
expression for the high temperature conduction of a SET valid for 
arbitrary tunneling conductance.

In the region of weak tunneling, $g<1$, the quantity $u$ becomes 
small at high temperatures and we 
may replace $\sigma(u)$ and $\tau(u)$ by
\begin{equation}
\sigma(0)=3 \zeta(3), \qquad
\tau(0)= \pi^4/10.
\end{equation}
This gives for the conductance 
of a weakly conducting SET
\begin{eqnarray}
\!\! G
&=&
\frac{1}{G_1^{-1}+G_2^{-1}}
\bigg\{
1-\frac{\beta E_c}{3}
                                  \nonumber\\
&&                     
\qquad \quad
+\left[ \frac{1}{15}
+g \frac{3 \zeta(3)}{2\pi^4}\right] (\beta E_c)^2
+{\cal O}(\beta E_c)^3
\bigg\},
\end{eqnarray}
in accordance with earlier work \cite{earlierwork}.
In the region of strong tunneling the quantity $u$ is typically large 
even for the highest temperatures explored experimentally and the full 
expression $(\ref{conductance})$, $(\ref{partition})$ must be used.

\begin{figure}
\begin{center}
\leavevmode
\epsfxsize=0.45 \textwidth
\epsfbox{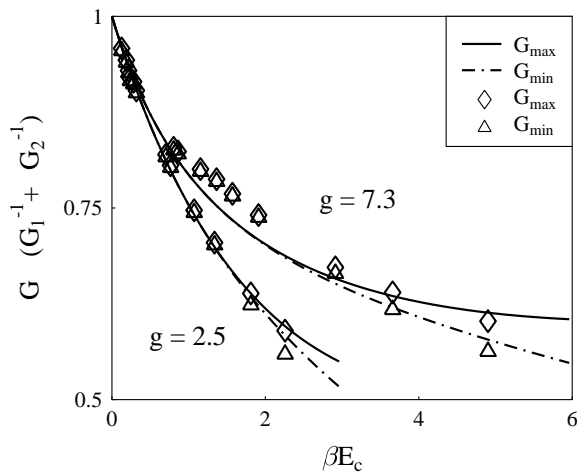}
\end{center}
\caption{Maximum and minimum linear conductance in dependence
on dimensionless temperature for two dimensionless parallel
conductances $\alpha=2.5$ and $7.3$ compared with experimental data 
(symbols) by Joyez et al. \protect\cite{Joyez}.}
\label{fig:fig2}
\end{figure}

We have compared our findings with recent experimental data by
Joyez et al.\ \cite{Joyez} for transistors with $g=0.6$, $2.5$ and $7.3$.
As seen from Fig.\ \ref{fig:fig2} the theory describes the high
temperature behavior of all junctions (results for $g=0.6$ are not 
shown) down to temperatures where the current starts to modulate 
with the gate voltage. The parameters have {\it not} been adjusted to improve
the fit but coincide with the values given in \cite{Joyez}. The small
deviations between theory and experiment for $g=7.3$ near $\beta
E_c=1$ may arise from experimental uncertainties in $\beta E_c$. 
We mention
that the temperature dependence of the conductance of the highly 
conducting SET ($g=7.3$) is not within reach of
previous theoretical predictions. The results obtained thus present
substantial progress and should be useful for experimental studies
of even larger tunneling conductances since the predictions remain valid
for arbitrary values of $g$.

The authors would like to thank Michel Devoret, Daniel Esteve, and 
Philippe Joyez for
valuable discussions. Financial support was provided by the Deutsche
Forschungsgemeinschaft (DFG) and the Deutscher Akademischer 
Austauschdienst (DAAD).

\end{multicols}
\end{document}